\documentclass[traditabstract,longauth]{aa}  
\usepackage{epsfig}
\usepackage{natbib}
\usepackage{txfonts}
\usepackage{graphicx}

\def\m2s2{\hbox{\,m$^{2}$\,s$^{-2}$}} 

\def\chisq{\mbox{$\chi^2$}\,}

\begin{document}

\title{The secondary eclipse of CoRoT-1b \thanks{Based on observations obtained with CoRoT, a space project operated by the French Space Agency, CNES, with participation of the Science Programme of ESA, ESTEC/RSSD, Austria, Belgium, Brazil, Germany and Spain.} }

\author{Alonso, R. \inst{1,2}
\and Alapini, A. \inst{3}
\and Aigrain, S. \inst{3}
\and Auvergne, M. \inst{4}
\and Baglin, A. \inst{4}
\and Barbieri, M. \inst{1}
\and Barge, P. \inst{1}
\and Bonomo, A.S. \inst{1}
\and Bord\'e, P. \inst{5}
\and Bouchy, F. \inst{6}
\and Chaintreuil, S. \inst{4}
\and De la Reza, R. \inst{7}
\and Deeg, H.J. \inst{8}
\and Deleuil, M. \inst{1}
\and Dvorak, R. \inst{9}
\and Erikson, A. \inst{10}
\and Fridlund, M. \inst{11}
\and Fialho, F. \inst{4}
\and Gondoin, P. \inst{11}
\and Guillot, T. \inst{12}
\and Hatzes, A. \inst{13}
\and Jorda, L. \inst{1}
\and Lammer, H. \inst{14}
\and L\'eger, A. \inst{5}
\and Llebaria, A. \inst{1}
\and Magain, P. \inst{15}
\and Mazeh, T. \inst{16}
\and Moutou, C. \inst{1}
\and Ollivier, M. \inst{5}
\and P\"atzold, M. \inst{17}
\and Pont, F. \inst{3}
\and Queloz, D. \inst{2}
\and Rauer, H. \inst{9,18}
\and Rouan, D. \inst{3}
\and Schneider, J. \inst{19}
\and Wuchterl, G. \inst{12}
}

\offprints{\email{roi.alonso@oamp.fr}}

\institute{Laboratoire d'Astrophysique de Marseille, UMR 6110, Technopole de Marseille-Etoile,F-13388 Marseille cedex 13, France
\and
Observatoire de Gen\`eve, Universit\'e de Gen\`eve, 51 Ch. des Maillettes, 1290 Sauverny, Switzerland
\and
School of Physics, University of Exeter, Stocker Road, Exeter EX4 4QL, United Kingdom
\and
LESIA, CNRS UMR 8109, Observatoire de Paris, 5 place J. Janssen, 92195 Meudon, France
\and
IAS, UMR 8617 CNRS, bat 121, Universite Paris-Sud, F-91405 Orsay, France
\and
Observatoire de Haute-Provence, 04870 St Michel l'Observatoire, France
\and
Observat\'orio Nacional, Rio de Janeiro, RJ, Brazil
\and
Instituto de Astrof\'\i sica de Canarias, E-38205 La Laguna, Spain
\and
Institute for Astronomy, University of Vienna, T\"urkenschanzstrasse 17, 1180 Vienna, Austria
\and
Institute of Planetary Research, DLR, Rutherfordstr. 2, 12489 Berlin, Germany
\and
Research and Scientific Support Department, European Space Agency, ESTEC, 2200 Noordwijk, The Netherlands 
\and
Observatoire de la C\^ote d'Azur, Laboratoire Cassiop\'ee, CNRS UMR 6202, BP 4229, 06304 Nice Cedex 4, France
\and
Th\"uringer Landessternwarte Tautenburg, Sternwarte 5, 07778 Tautenburg, Germany
\and
Space Research Institute, Austrian Academy of Sciences, Schmiedlstrasse 6, 8042 Graz, Austria
\and
Institut d'Astrophysique et de G\'eophysique, Universit\'e de  Li\`ege, All\'ee du 6 ao\^ut 17, Sart Tilman, Li\`ege 1, Belgium
\and
School of Physics and Astronomy, R. and B. Sackler Faculty of Exact Sciences, Tel Aviv University, Tel Aviv 69978, Israel
\and 
Rheinisches Institut f\"ur Umweltforschung, Universit\"at zu K\"oln, Abt. Planetenforschung, Aachener Str. 209, 50931 K\"oln, Germany
\and
Center for Astronomy and Astrophysics, TU Berlin, Hardenbergstr. 36, D-10623 Berlin, Germany
\and
LUTH, Observatoire de Paris-Meudon, 5 place J. Janssen, 92195 Meudon, France
}

\date{Received 18 March 2009 / Accepted 7 July 2009}

\abstract{The transiting planet CoRoT-1b is thought to belong to the $pM$-class of planets, in which the thermal emission dominates in the optical wavelengths. We present a detection of its secondary eclipse in the CoRoT white channel data, whose response function goes from $\sim$400 to $\sim$1000~nm. We used two different filtering approaches, and several methods to evaluate the significance of a detection of the secondary eclipse. We detect a secondary eclipse centered within 20~min at the expected times for a circular orbit, with a depth of 0.016$\pm$0.006\%. The center of the eclipse is translated in a 1-$\sigma$ upper limit to the planet's eccentricity of $e\cos\omega<$0.014. Under the assumption of a zero Bond Albedo and blackbody emission from the planet, it corresponds to a T$_{CoRoT}$=2330$^{+120}_{-140}$~K. We provide the equilibrium temperatures of the planet as a function of the amount of reflected light. If the planet is in thermal equilibrium with the incident flux from the star, our results imply an inefficient transport mechanism of the flux from the day to the night sides.}


\titlerunning{The secondary eclipse of CoRoT-1b}

\authorrunning{Alonso et al.}

\maketitle

\section{Introduction}
\label{sec:intro}

When a Hot Jupiter transits its host star, unless the eccentricity is
anomalously large, it will also produce secondary eclipses, sometimes also called
occultations, from which precious information about the atmospheres of
these intriguing objects can be inferred.  During the last four years,
we have witnessed the first detections of thermal emission from
several Hot Jupiters at different band passes in the infrared, most of
them from space (\citealt{char05,dem05}). These pioneering studies
have revealed some features of their atmospheres, such as thermal
inversions at hight atmospheric altitudes (\citealt{knu08,knu09}).

The different opacity of the planetary atmosphere, especially due to the abundances of TiO and VO, can make the spectrum emitted by the planet very different, what led \cite{fort}
to propose two classes of exoplanets, $pM$ and $pL$. Those receiving
more incident flux from their star, the $pM$ class, would exhibit
thermal inversions of their stratospheres, and as a consequence the
depths of the secondary eclipses in the optical and near infrared
bands will be bigger than expected for a $pL$ planet. These authors,
as well as \cite{mer}, argue that in the red parts of the optical
spectrum, the thermal emission from these objects is much more
important than the contribution of the reflected light, as theoretical
models are in agreement with a very low albedo $A$ for $pM$ planets (\citealt{sud00,burrows,hood}).

The first transiting planet detected from space,
CoRoT-1b~\citep{bar08}, completes an orbit around its G0V star
every 1.5~d. Its incident flux puts this planet in the $pM$
category. In this paper, we focus on the detection of its secondary
eclipse. We describe the data set and the preparation of the light
curve for the search for the secondary in Section~\ref{sec:obs}, and
the different techniques used to evaluate the significance and depths
of the secondary in Section~\ref{sec:ana}. Finally, in
Section~\ref{sec:dis}, we discuss the physical implications of our
results.

\section{Observations}
\label{sec:obs}

CoRoT-1b was observed during the first observing run of CoRoT,
attaining a total duration of 52.7~d. In the present study, the
photometry was performed using the latest version of the pipeline, 
which uses the information about the
instrument's Point Spread Function and the centroids of the stars
measured in the asteroseismic channel to correct for the effects of
the satellite jitter in the white light curve. The effect of the Earth
eclipses, in which a thermal shock is translated in a bigger jitter
and thus less flux inside the photometric apertures, is greatly
improved in this new version of the pipeline.

\begin{figure}[!th]
\begin{center}
\epsfig{file=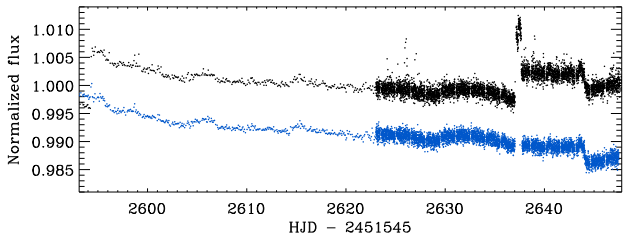,width=9cm}
\caption{Light curve of CoRoT-1b prepared for secondary eclipse
  search (the transits have been removed). For displaying purposes, the data were combined in 10-points
  bins. The black points show the pipeline version of the curve, while the blue points show the curve corrected for the orbital residuals, hot pixels and remaining outliers, as described in the text. }
\label{fig:fig1}
\end{center}
\end{figure}

\begin{figure}[!th]
\begin{center}
\epsfig{file=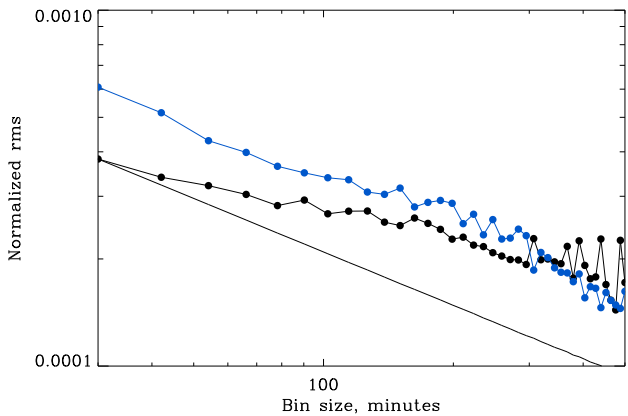,width=9cm}
\caption{The standard deviation of the light curve after removing a median filtered version with a 12-h window, and after combining it in bins of different sizes. Black dots are for the 32-s part of the light curve, and blue dots for the 512-s part. The solid line follows the theoretical diminution of the dispersion from the 30 min point in the 32-s part, if the noise were white.}
\label{fig:fig_rms}
\end{center}
\end{figure}

The light curve was sampled every 512~s for the first 28~days, and
then changed to 32~s until the end of the run. During the 512~s part,
the dispersion of the normalized data is 0.00079, and 0.0019 in the
32~s section. The pipeline version of the light curve is plotted in 
Fig.~\ref{fig:fig1}.
To evaluate the red noise content in our data (e.g. \citealt{pont}), we computed the standard deviation of the light curve (filtered with a median filter with a window of 12 h) with different bin sizes, ranging from 30 to 500 min. The result is plotted in Fig.~\ref{fig:fig_rms}; the 32-s part of the curve shows less noise than the 512-s part below bins of $\sim$150 min, while it achieves similar levels as the 512-s part above that size. This reveals the presence of uncorrected low frequencies at levels of $<$10$^{-4}$ in units of normalized flux, and of uncorrected signals of about 2$\times$10$^{-4}$ with periods between 30 and 150 min in the 512-s part of the data.

\section{Analysis of the CoRoT light curve}
\label{sec:ana}

We present two different analysis of the data in order to search and measure the secondary eclipse of CoRoT-1b. They differ substantially from the preparation of the light curve to the estimation of the significance of the secondary eclipse detection, and we considered both in order to reinforce or invalidate a detection of the secondary eclipse.

\subsection{First analysis: filtering, local polynomial normalization and trapezoid fit}

Two jumps in the data, due to hot pixels, were corrected by estimating
the median fluxes before and after the jump, and we rejected the
points belonging to the region between CoRoT date (HJD - 2451545)
2637.1 and 2637.75, as they were affected by the stabilization of the
hot pixel to its normal levels.
 
\begin{figure}[!th]
\begin{center}
\epsfig{file=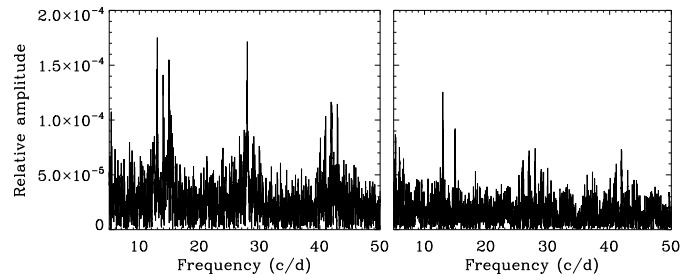,width=9cm}
\caption{Amplitude spectrum of the light curve of CoRoT-Exo-1, before
  (left) and after (right) the correction for the orbital residuals
  described in the text.}
\label{fig:figp04}
\end{center}
\end{figure}

The signal we want to measure is at the 10$^{-4}$ level, and thus we
inspected the level of noise in the light curve by producing its
amplitude spectrum using Period04~\citep{lenz05}. In the left panel of
Fig.~\ref{fig:figp04}, some peaks at the orbital frequency and its
daily aliases remain at a level of $1.5 \times 10^{-4}$. This
motivated us to try to improve the filtering at the orbital period. We
estimated the signal at each satellite's orbit $i$ from the signal in
the 30 orbits before and after the orbit $i$, as described in
\cite{alo08a}. Some remaining outliers were located by subtracting the
low frequencies in the curve using a moving median filter with a
window of about 1~d, and flagging iteratively the points at more than
3 times the standard deviation calculated in a robust way. Finally,
these points were replaced by interpolations to their closest
points. In total, these interpolated points represent 3.0\% of the
original data. The amplitude spectrum of the corrected curve is
plotted in the right panel of Fig.~\ref{fig:figp04}, where the
amplitudes of the peaks at the orbital frequency and its aliases and
overtones have clearly been reduced.

To avoid possible effects of unequal number of points at different
orbital phases due to the phases at the beginning and end of the
observations, we cut the parts of the data before the first recorded
transit and after the last one. As we are interested in the secondary
eclipse, we also cut the transits in the curve. The light curve
prepared for the search for the secondary eclipse is plotted in
Fig.~\ref{fig:fig1}.

\begin{figure}[!th]
\begin{center}
\epsfig{file=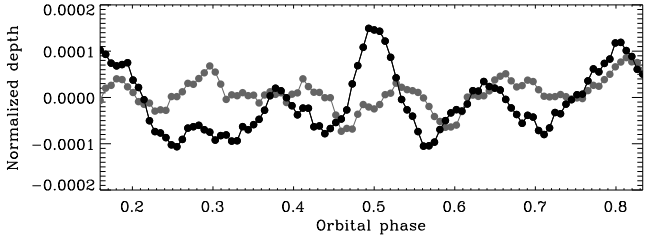,width=9cm}
\caption{Depth of a trapezoid with the shape and duration of a
  transit, as a function of the planet's orbital phase. The maximum is
  centered in phase 0.5, corresponding to the secondary eclipse. In
  grey, the result in a light curve where the secondary eclipse signal
  has been diluted (see text for details).}
\label{fig:pha1}
\end{center}
\end{figure}

\begin{figure}[!th]
\begin{center}
  \epsfig{file=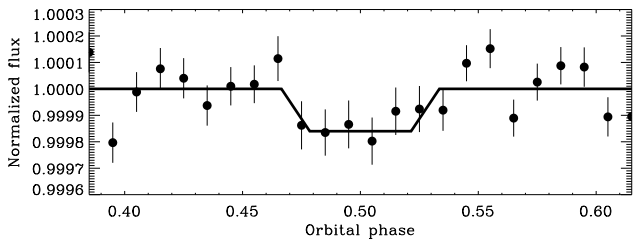,width=9cm}
  \caption{Phase folded curve of CoRoT-1b during the phases of
    secondary eclipse. The data have been binned in 0.01 in phase
    ($\sim$22~min).}
\label{fig:fig6}
\end{center}
\end{figure}

\begin{figure}[!th]
\begin{center}
  \epsfig{file=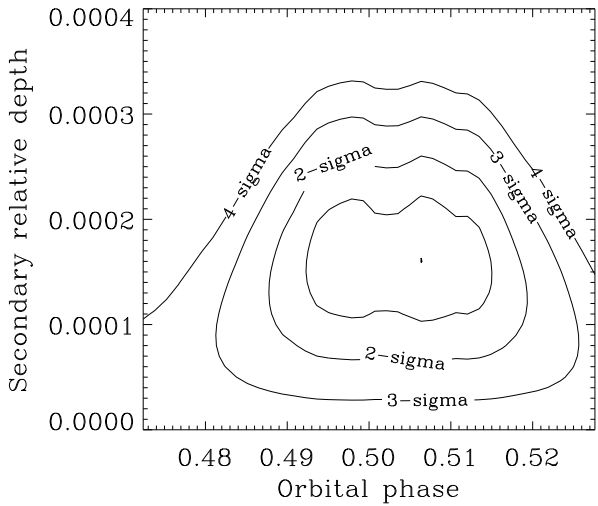,width=9cm}
  \caption{The \chisq space for different centers and depths of the
    secondary eclipse, and the different $\sigma$ confidence limits.}
\label{fig:fig8}
\end{center}
\end{figure}

\begin{figure}[!th]
\begin{center}
  \epsfig{file=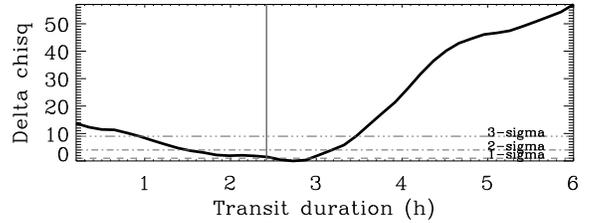,width=9cm}
  \caption{The \chisq for the duration of the secondary eclipse, and
    the 1,2 and 3$\sigma$ confidence limits. The vertical line shows
    the total duration of a transit. The detected signal has the same
    duration as the transit at 1-$sigma$ level.}
  \label{fig:fig7}
\end{center}
\end{figure}

For the search of the secondary eclipse, we followed the method
described in \cite{alo08b}. Basically, it consists in evaluating the
depths of fits to trapezoids with the same overall shape as the
planet's transit, at different phases of the orbital period. At each
test phase $\phi_i$, the low frequencies are removed by performing, at
each planetary orbit, linear fits to the regions
$[\phi_i-\phi_a-\phi_b,\phi_i-\phi_a)\bigcup(\phi_i+\phi_a,\phi_i+\phi_a+\phi_b]$, where $\phi_a$
and $\phi_b$ were chosen in order not to include points inside a
transit duration, but at the same time allowing enough data points to
enter in the fit. We obtained a good compromise using the values
$\phi_a$= 0.04 and $\phi_b$=0.12. Once all the planetary orbits were
normalized this way, we binned the data with a size of 0.001 in phase,
and fitted the depth of a trapezoid using a Levenberg/Marquardt
algorithm (\citealt{lev,mar}), fixing the rest of the trapezoid's
parameters to the values of the transit. The final depth of the fits
as a function of the orbital phase is plotted in Fig.~\ref{fig:pha1},
where the maximum is well centered at the expected orbital phase
0.5. We can evaluate the significance of this detection by computing
the dispersion of the fitted depths in the parts of the phase diagram
not affected by the inclusion of the secondary eclipse at phase 0.5 in
the regions where the fits used to normalize the transits were
computed, i.e.,
$\phi\in[0,0.5-\phi_a-\phi_b)\bigcup(0.5+\phi_a+\phi_b,1]$. The
significance of the detection calculated this way results in
2.1-$\sigma$. The phase folded light curve
around the secondary phase and the best fit trapezoid (with duration
and shape fixed to that of the transits) are shown in
Fig.~\ref{fig:fig6}.

We performed the same technique in a light curve where we diluted the
signal of the secondary eclipse. To do so, we subtracted from the
light curve a version of it, filtered with a moving median using a
window of $\sim$1~d in duration. We shuffled randomly the residuals,
and we added back the subtracted filtered curve. The resulting depth
vs. phase diagram using the method described above in this curve is
plotted as a grey line in the Fig.~\ref{fig:pha1}. If we take the
dispersion of this measured depth (3.29$\times$10$^{-5}$) as the
precision in the measurement of the secondary eclipse, then the
significance of the signal at phase 0.5 is 4.5-$\sigma$.

Additionally, three different methods to evaluate the depth and
significance of the secondary eclipse signal were tested. In the first
method, we removed the best fit trapezoid to the data (where the only
fitted parameter was the depth, the rest was fixed to the values of
the transit), shifted circularly the residuals, reinserted the signal,
and re-evaluated the fitted depth. The final depth thus takes into
account the effect of red noise in the data, and it is of
0.014$\pm$0.002\%. The second method consisted in fitting two
gaussians to 1) the distribution of points inside total eclipse and 2)
a subset of the points outside the eclipse with the same number of
points as in 1), and compare their fitted centers. This fit was
performed to 500 subsets of 2), with a randomly chosen starting point
and varying the size of the bins in the distribution between 0.005\%
and 0.02\%. The result is of 0.021$\pm$0.003\%. In the third method, we explored
the \chisq distribution in a grid of centers (from -60 to +60 minutes
from the expected center) and depths (from 0.01\% to 0.04\%) of the
secondary eclipse. The other parameters of the trapezoid were fixed to the values of the transit. The minimum \chisq and the 1,2 and 3-$\sigma$
confidence levels are presented in Fig.~\ref{fig:fig8}, and the best
fitted depth using this method is 0.016$\pm$0.006\%.  We show the \chisq map of the duration
of the trapezoid and the 1,2,3-$\sigma$ confidence levels in
Fig.~\ref{fig:fig7}. The duration of the secondary eclipse is, as
expected, compatible at a $<$2-$\sigma$ level with the duration of the
transits.

\subsection{Second analysis: using the IRF filter}

The analysis presented so far shows encouraging evidence for the
detection of a secondary eclipse at the expected phase and duration
for a circular or almost circular orbit. However, the detection is
unavoidably tentative given the extremely shallow nature of the
signal. Each step of the analysis involved a number of free
parameters, from the hot pixel and satellite orbit residual
corrections to the individual corrections applied for stellar
variability local to each putative secondary location. 

In an effort to reinforce or invalidate the detection, we carried out
a separate secondary eclipse search using different pre-processing and
eclipse detection methods which were designed to minimize the number
of free parameters. The regions around the hot pixel events were
simply clipped out, no correction for orbital residuals was applied,
and we used the iterative reconstruction filter (IRF) of \citet{aa09},
which has only two free parameters, to isolate signal at the planet's
orbital period from other signals including stellar variability.
 
The starting point of this analysis was the same light curve, as
described in Section~\ref{sec:obs}. To circumvent the issue of varying
data weights associated with different time sampling, the oversampled
section of the light curve was rebinned to 512s sampling. Outliers
were then identified and clipped out using a moving median filter (see
\citealt{apf09} for details). Finally, we also discarded two segments
of the light curve, in the CoRoT date ranges 2594.25--2594.45 and
2637.10--2637.70), corresponding to the two hot pixel events visible
in Fig.~\ref{fig:fig1}. This last
step was necessary as the IRF cannot remove the sharp flux variations
associated with hot pixels without affecting the transit signal.

The resulting time-series was then fed into the IRF. A full
description of this filter is given in \citet{aa09}, but we repeat the
basic principles of the method here for completeness. The IRF treats
the light curve $\{Y(i)\}$ as $Y(i)=F(i)\,A(i)+R(i)$, where $\{A(i)\}$
represents the signal at the period of the planet, which is a
multiplicative term applied to the intrinsic stellar flux $\{F(i)\}$,
and $\{R(i)\}$ represents observational noise. A first estimate of
$\{A(i)\}$ is obtained by folding the light curve at the period of the
transit and smoothing it using a smoothing length of 0.0006 in phase,
and is divided into the original light curve. This is then run through
an iterative non-linear filter \citep{ai04} which preserves signal at
frequencies longer than 0.5 days to give an estimate of the stellar
signal $\{F(i)\}$, which is assumed to be primarily concentrated on
relatively long time-scales. This signal in turn is removed from the
original light curve and the process is iterated until the the
dispersion of the residuals remains below $10^{-4}$ for 3 consecutive
iterations, which occurs after 3 iterations in this case.

\begin{figure}[!th]
\begin{center}
  \epsfig{file=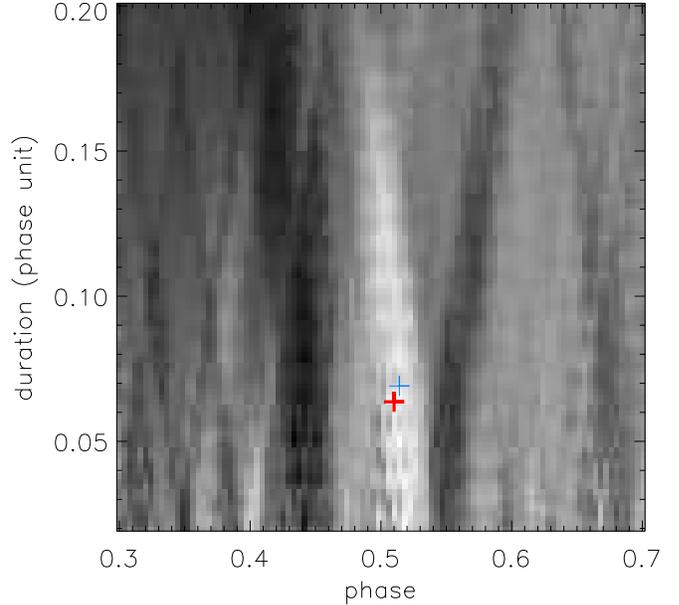,width=9cm}
  \caption{Detection significance map. The red thick cross is the
    secondary eclipse duration and phase with the largest detection
    significance. The blue thin cross is the largest detection
    significance for a secondary eclipse of the same duration as the
    primary transit.}
\label{fig:fig_s4_1}
\end{center}
\end{figure}

\begin{figure}[!th]
\begin{center}
  \epsfig{file=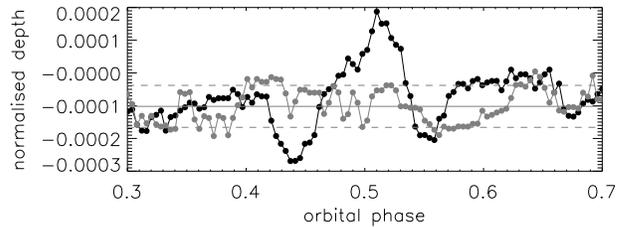,width=9cm}
  \caption{Depth of a box as a function of the planetÕs orbital phase (black curve) with the total eclipse duration (0.0697 in phase units) providing the highest detection significance. The largest significance is found for a secondary eclipse depth of 0.019\% at an orbital phase of 0.510. The grey dashed lines show the scatter ($\pm$0.006\%) of depths measured in a light curve freed from the secondary eclipse with shuffled residuals (grey curve). This scatter is an estimate of the uncertainty on the depth measurements.}
\label{fig:fig_s4_2}
\end{center}
\end{figure}

The free parameters of the IRF are the smoothing lengths used when
estimating $\{A(i)\}$ and $\{F(i)\}$. The former represents a
compromise between reducing the noise and blurring out potential sharp
features associated with the planet, and the latter between removing
the stellar signal without affecting the planetary signal. The values
chosen here were adopted by trial and error and gave the best results
when evaluating the performace of the IRF for transit reconstruction
purposes \citep{aa09}.

We searched the IRF-filtered light curve for secondary eclipses using
a very simple sliding box algorithm, where the model is taken to be 1
outside the eclipse and the in-eclipse level is set to the median of
the in-eclipse points (the light curve was previously normalised by
dividing it be the median of the out-of-transit points). This is
therefore a 2-parameter model, and the phase and duration of the
eclipse were varied over a $100\times100$ grid ranging from 0.3 to
0.7 in phase and from 0.02 to 0.2 in duration, where the duration is
measured in phase units (corresponding to 0.7 to 7.2 hours). We then
evaluate the significance of the detection, in arbitrary units, by 
dividing the eclipse depth by $\sigma/\sqrt(N)$, where sigma is the scatter of the points - outside the primary transit - in the IRF-filtered light curve and $N$ is the number of point used to calculate the level inside the secondary eclipse\footnote{The significance can be estimated as S=$depth/\sigma \times \sqrt{N}$, and in this case the dispersion $\sigma$, and the total number of points N remain constant.}. The
resulting confidence map is shown in Fig.~\ref{fig:fig_s4_1}, where
the phase and duration giving the best significance are marked with a
red cross. The best duration corresponds almost exactly to the full
duration of the transit (which we measure as 0.0688 in phase units
from a trapezoidal fit), and the best phase is 0.510, in good
agreement with the results of Fig.~\ref{fig:fig8}. The diagonal
patterns in the significance map are due to individual features in
the folded, filtered light curve, which influence a wider range of
trial phases at longer durations.

Fig.~\ref{fig:fig_s4_2} shows the eclipse depth as a function of
orbital phase at the best duration. The best-fit depth is 0.019\%,
consistent with the results presented above. 
One should bear in mind when comparing the different results that the eclipse depth was measured here relative to the
overall median flux rather than to a local estimate of the
out-of-eclipse flux.

To calculate the significance of the detection, we measured the depth and associated uncertainty and divided the former by the latter. In this case, we calculated the uncertainty as the dispersion (estimated as 1.48$\times$MAD) around the median level of the light curve outside the primary transit, and divided this value by the square root of the number of points inside the secondary eclipse. This results in an uncertainty of 0.005\%, i.e.\ a 3.5 $\sigma$ detection. We also applied the three other methods already presented in Section 3.1. All of these give an uncertainty estimate of 0.006\%, which we adopt as the final uncertainty on the eclipse depth derived from the IRF.

The fact that we arrived at consistent estimates of the secondary eclipse depth and significance using different approaches to filter the light curve and evaluate the uncertainty on the depth lends confidence to the detection. Our final adopted values for the depth is 0.016$\pm$0.006\%, consistent with all our estimates within one sigma. The eclipse center occurs at the expected phase for a circular orbit, with an uncertainty of about 20 min.

\section{Discussion}
\label{sec:dis}

We have shown that a decrease in the flux of the light curve of CoRoT-1b at the phases of secondary eclipse is detected at more than 3-$\sigma$ level. Its duration is, within 1-$\sigma$, the same as the duration of the transits, and its shallow depth is of only 0.016$\pm$0.006\%. We thus interpret this signal as the secondary eclipse, detected at the optical part of the spectrum. 
 
If we were to explain the secondary eclipse detection as reflected light, it would imply a geometric albedo of $A_g$=0.20$\pm$0.08, which is a bigger value than several upper limits provided by other works in different exoplanets (\citealt{andrew,rowe}). Furthermore, theoretical models predict a strong optical absorption in the Hot Jupiters' atmospheres (e.g.~\citealt{sud00,burrows,hood}), and for the $pM$-class of planets, thermal emission dominates by more than an order of magnitude the reflected light. Consequently, we suspect that the secondary eclipse signature is not produced entirely by reflected light, but most probably by a combination of thermal emission and some reflected light, as in the case of CoRoT-2b~\citep{alo09b}.

The CoRoT bi-prism allows to recover chromatic information from the signal. In this case, the difference between the significance of the secondary signal in the blue and the red channels might clarify if our detected signal is dominated by reflected light or by thermal emission in the optical, as we assumed given the arguments above. In the first case, the secondary should be detected mostly in the blue channel, while in the case of thermal emission the red channel would carry most of the signal. We checked the colors in the  CoRoT-1 aperture, but unfortunately the residuals of the jitter correction and the noisier individual channels did not allow us to conclude on this subject. \footnote{After submission of the original manuscript of this paper, we became aware of an independent analysis of the red CoRoT channel performed by \cite{sne09}, that allowed these authors to detect a 0.0126$\pm$0.0033\% secondary eclipse. This result points towards a small fraction of reflected light in the white curve, as if thermal emission were the only source of received flux we would expect a deeper secondary eclipse in the red channel than in the white channel.}

The equilibrium temperature can be calculated as 
\begin{equation}
T_{eq}=T_\star(R_\star/a)^{1/2}[f(1-A_B)]^{1/4}
\label{eq} 
\end{equation}
which depends on the Bond albedo $A_B$ and the re-distribution factor $f$ which accounts for the efficiency of the transport of energy from the day to the night side of the planet. $f$ can vary between $1/4$ for an extremely efficient redistribution (isothermal emission at every location of the planet) and higher values for an inefficient redistribution, implying big differences in the temperatures at the day/night sides of the planet.

\begin{figure}[!th]
\begin{center}
  \epsfig{file=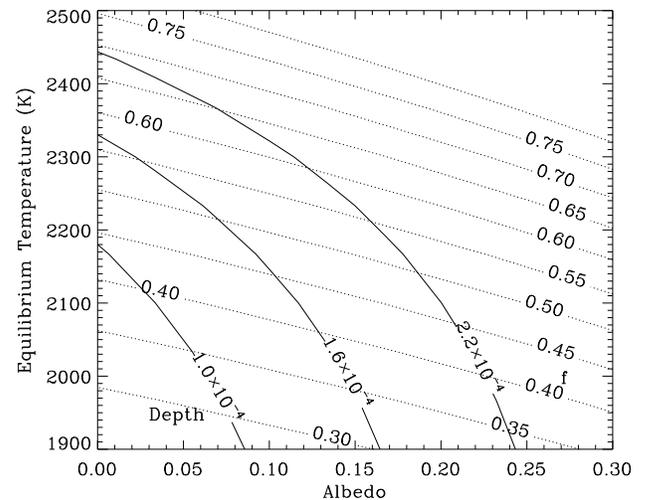,width=9cm}
  \caption{The implied equilibrium temperatures from the secondary eclipse measurement, as a function of the amount of reflected light. The best fitted depth and the one sigma limits are plotted as the solid lines. Different values of the re-distribution factors as defined by Eq.~\ref{eq} are plotted as dotted lines. Zero albedo solutions imply high values of the re-distribution factor.}
\label{fig_teff}
\end{center}
\end{figure}

To translate the measured depth of the secondary eclipse into brightness temperatures, we used a theoretical model of a G0V star from the Pickles library of stellar models \citep{pickles}, calibrated in order to produce the same integrated flux as a Planck black-body spectrum with a T$_{\rm{eff}}$=5950~K. Under the assumption that the planetary emission is well reproduced by a black-body spectrum, and as we know the ratio R$_p$/R$_\star$ from the transits, we can calculate the temperatures that produce secondary eclipse depths compatible with our result, using the CoRoT response function given in \cite{auv09}. The brightness temperature calculated this way resulted in T$_{\rm{CoRoT}}$=2330$^{+120}_{-140}$~K.  If we further assume that the planet is in thermal equilibrium and a zero Bond Albedo, this temperature favors high values of the re-distribution factor $f$=0.57$^{+0.10}_{-0.13}$. A diagram showing the implied equilibrium temperatures as a function of the albedo, where we have assumed for simplicity that all the reflected light is detected in the white channel of CoRoT (i.e., that $A_g$=$A_B$, as in Alonso et al. 09b) is shown in Fig.~\ref{fig_teff}.

Non-zero Bond Albedos, where some of the detected flux is reflected light from the star, would imply smaller values of $f$, whilst solutions with A$>$0.30 are not compatible with the measured secondary eclipse. All the solutions with equilibrium temperatures higher than 1900~K are for $f$ values bigger than 1/4. Thus, the secondary eclipse detection favors inefficient redistributions of the incident flux from the day to the night side. 

One may wonder if the $k=R_{pl}/R_\star$ obtained from the transit fit should be different because of the lack of inclusion of the emitted light from the planet in the model. Even for an extreme case of $f=1/4$, the correction to be applied on the $k$ is a factor of 3 below the uncertainty given in \cite{bar08}, and thus we do not consider it necessary. 
 
The measured center of the secondary eclipse, with an uncertainty of 20~min, can be used to constrain the $e\cos\omega<$0.014 at a 1-$\sigma$ level.

Our measured brightness temperature can be used to predict eclipse depths in the $K_s$ and $z$ bands of 0.25\% and 0.014\% respectively. In these bands, ground-based observations have recently been successful in the detections of secondary eclipses of exoplanets (\citealt{sing09,mooij09}), and for the case of CoRoT-1b, the observations might reveal departures from the black-body assumption such as the thermal inversions observed in several planets (e.g. \citealt{knu08,knu09}). \footnote{After submission of this work, \cite{gil09} reported a detection in the $K_s$-band of a 0.28$^{+0.04}_{-0.07}$\% eclipse, in good agreement with the extrapolation of our results.}

\begin{acknowledgements}
R.A acknowledges support by the grant CNES-COROT-070879. A.H. acknowledges the support of DLR grants 50OW0204, 50OW0603, and 50QP0701. H.J.D. acknowledges support by grants ESP2004-03855-C03-03 and
ESP2007-65480-C02-02 of the Spanish Education and Science ministry.

\end{acknowledgements}

\end{document}